\documentclass[twocolumn,showpacs,preprintnumbers,amsmath,amssymb]{revtex4}
\usepackage{graphicx}
\usepackage{dcolumn}
\usepackage{bm}
\def\bea {\begin{eqnarray}}
\def\eea {\end{eqnarray}}

\def\be {\begin{equation}}
\def\ee {\end{equation}}

\begin{document}

\title{Soft gluon multiplicity distribution revisited }

\author{Santosh K Das and Jan-e Alam}

\medskip

\affiliation{Variable Energy Cyclotron Centre, 1/AF, Bidhan Nagar , 
Kolkata - 700064}

\date{\today}
\begin{abstract}
In this paper the soft gluon radiation from partonic interaction of the 
type: $2\,\rightarrow\, 2\,$ + gluon has been revisited and a correction term 
to the widely used Gunion-Bertsch (GB) formula  is obtained.
\end{abstract}

\maketitle

\begin{figure}[h]
\begin{center}
\includegraphics[scale=0.43]{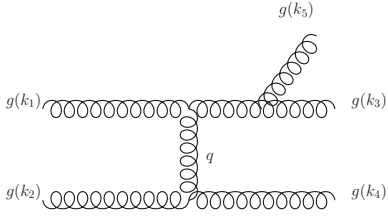}
\caption{A typical Feynman diagram for the process: $gg\rightarrow ggg$}
\label{fig1}
\end{center}
\end{figure}

The energy loss of high energy partons propagating 
through a thermalized system of quarks and gluons created in heavy ion 
collisions at relativistic energies has been measured through nuclear 
suppression factors at Relativistic Heavy Ion Collider (RHIC) 
energy~\cite{rhic}. The two most important 
mechanisms for the energy loss are radiative and collisional processes. 
Therefore, it is very important to understand and theoretically improve 
the  calculations of partonic energy loss in thermal medium.
Generically the radiative energy loss can be written as 
$2 \rightarrow 2 + g$ process, here we consider the 
process $gg\rightarrow gg + g$,  and the results
for other process can be obtained from it in a straight forward way. 
The typical Feynman diagram for $gg\rightarrow  gg+g$ 
is shown in Fig.~\ref{fig1}.
The  momentum $k_5$ of the radiative gluon (Fig.~\ref{fig1}) 
is taken to be a soft 
radiation around zero rapidity in 
the centre of momentum frame. The invariant amplitude for this
process in medium is given by (see the appendix):
\begin{eqnarray}
\left| M_{gg \rightarrow ggg} \right|^2=\left(\frac{4g^4N_c^2}{N_c^2-1×}
\frac{s^2}{(q_\perp^2+m_D^2)^2×}\right) \nonumber \\
\left(\frac{4g^2N_cq_\perp^2}{k_\perp^2\lbrack(k_\perp-q_\perp)^2+m_D^2\rbrack} \right) \nonumber \\
+\frac{16g^6N_c^3}{N_c^2-1×}\frac{q_\perp^2}{k_\perp^2\lbrack(k_\perp-q_\perp)^2+m_D^2\rbrack}
\label{eq1}
\end{eqnarray}
where $k_\perp$ and $q_\perp$ are the perpendicular component of $k_5$ and that 
of the momentum transfer in the centre of momentum frame respectively, $m_D$ is
the Debye mass, $N_c$ is the number of colour degrees of freedom. 
The first parenthesis in the first term in Eq.~\ref{eq1} stands for the
square of the invariant amplitude for the process: 
$gg \rightarrow gg$
\be
\left| M_{gg \rightarrow gg} \right|^2=\left(\frac{4g^4N_c^2}{N_c^2-1×}\frac{s^2}{(q_\perp^2+m_D^2)^2×}\right)
\label{eq2}
\ee
and the second parenthesis in the first term represents the soft gluon 
emission spectrum~\cite{wong}. 
The second term of Eq.~\ref{eq1} is the correction to the  squared modulus 
of the matrix elements.  The first term of Eq.~\ref{eq1} is reported in 
Ref.~\cite{wong}. In the limit $m_D\rightarrow 0$ the first 
term  also reproduces the results obtained in Ref.~\cite{GB}.

The soft gluon multiplicity distribution at fixed $q_\perp$ 
is given by~\cite{GB,wong}:

\begin{figure}[h]
\begin{center}
\includegraphics[scale=0.43]{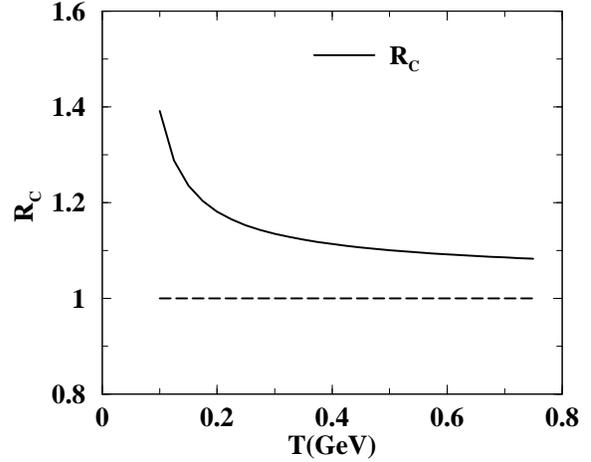}
\caption{The variation of $R_c$ (see Eq.~{\protect{\ref{eq6}}}) 
with temperature.
}
\label{fig2}
\end{center}
\end{figure}
\begin{eqnarray}
\frac{dn_g}{d\eta dk_\perp^2×}=\frac{C_A\alpha_s}{\pi^2} 
\left(\frac{q_\perp^2}{k_\perp^2\lbrack(k_\perp-q_\perp)^2+m_D^2\rbrack×} \right) + \nonumber \\
\frac{C_A\alpha_s}{\pi^2}
\left(\frac{q_\perp^2(q_\perp^2+
m_D^2)^2}{s^2k_\perp^2\lbrack(k_\perp-q_\perp)^2+m_D^2\rbrack×} \right)
\label{eq3}
\end{eqnarray}

In Eq.~\ref{eq3} the second term is the correction to the 
soft gluon multiplicity distribution.
Eq.~\ref{eq3} can be written as:
\begin{eqnarray}
\frac{dn_g}{d\eta dk_\perp^2×}=
\left[\frac{dn_g}{d\eta dk_\perp^2×}\right]_{GB}
\left(1+\frac{(q_\perp^2+m_D^2)^2}{s^2×}\right)
\label{eq4}
\end{eqnarray}
where
\begin{equation}
\left[\frac{dn_g}{d\eta dk_\perp^2×}\right]_{GB}=
\frac{C_A\alpha_s}{\pi^2} 
\frac{q_\perp^2}{k_\perp^2\lbrack(k_\perp-q_\perp)^2+m_D^2\rbrack×}
\label{eq5}
\end{equation}

To estimate the contributions from the correction term 
we consider the ratio, $R_c$ given by
\begin{equation}
R_c=\frac{\frac{dn_g}{d\eta dk_\perp^2×}}
{\left[\frac{dn_g}{d\eta dk_\perp^2×}\right]_{GB}}=
1+\frac{(q_\perp^2+m_D^2)^2}{s^2×}
\label{eq6}
\end{equation}
We evaluate $R_c$ by substituting $s=<s>=18T^2$, 
$m_D=\sqrt{4\pi\alpha_s(T)}\,T$ and  
$q_\perp^2=<q_\perp^2>$ which is calculated by using the
following relation:
\begin{equation}
<q_\perp^2>=\frac{\int dt\,t\,(d\sigma/dt)}{\int dt\,(d\sigma/dt)}
\label{eq7}
\end{equation} 
The lower and upper limits of the above integration are
$=m_D^2$ and $s/4$ respectively.
The variation of $R_c$ with $T$ is depicted in Fig.~\ref{fig2}.
The temperature dependence of the strong coupling $\alpha_s$ is taken from~\cite{zantow}.
It is observed that the correction to the gluon spectrum is appreciable for low
temperature domain. 

The similar correction in the soft gluon multiplicity distribution 
can be obtained for the processes 
$qg \rightarrow qgg$ and $qq \rightarrow qqg$. 
This result can also be used for radiative energy loss of heavy
quarks due to gluon emission.
The effects of quark mass can be taken 
into account by multiplying the emitted gluon 
distribution from massless quarks by a factor, $F^2$
which takes in to account the dead cone effects~\cite{DK}:
\be
F=\frac{k_\perp^2}{\omega^2\theta_0^2+k_\perp^2}
\label{eq8}
\ee
where $\theta_0=M/E$, $M$ is the mass and $E$ is the energy of the heavy
quarks. 
The results for the gluon spectrum emitted 
by a heavy quark of mass $M$  can be obtained
by multiplying the Eq.~\ref{eq4} by $F^2$.
%
In summary we have derived an expression for the soft gluon
multiplicity distribution from a process of the type:  $2\rightarrow 2+g$.
We observe that the corrections to the gluon spectrum obtained by~\cite{GB,wong}
is non-negligible in the low temperature region.
This result will be  useful in estimating the energy loss of high energy 
partons propagating through a thermalized system of quarks and gluons.

\section{Appendix}
In this appendix we outline the derivation of Eq.~\ref{eq1}.
The invariant amplitude for the process, $gg\rightarrow ggg$
can be written as~\cite{berends}:
\begin{eqnarray}
\left| M_{gg \rightarrow ggg} \right|^2=\frac{1}{2×}g^6\frac{N_c^3}{N_c^2-1×}\lbrack(12345)+
(12354)  \nonumber \\ +(12435)+(12453)+(12534)+(12543) \nonumber \\  +(13245)+(13254)+(13425)+(13524) \nonumber \\  
+(14235) +(14325)\rbrack \frac{N}{D}
\label{eq9}
\end{eqnarray}
where
\begin{eqnarray}
N=(k_1k_2)^4+(k_1k_3)^4+(k_1k_4)^4+(k_1k_5)^4  \nonumber  \\      +(k_2k_3)^4+ (k_2k_4)^4+ 
(k_2k_5)^4+(k_3k_4)^4     \nonumber  \\   +(k_3k_5)^4+(k_4k_5)^4,
\label{eq10}
\end{eqnarray}

\begin{eqnarray}
D=k_1.k_2k_1.k_3k_1.k_4k_1.k_5k_2.k_3k_2.k_4  \nonumber  \\  k_2.k_5k_3.k_4k_3.k_5k_4.k_5
\label{eq11}
\end{eqnarray}
and
\begin{eqnarray}
(ijklm)=(k_i.k_j)(k_j.k_k)(k_k.k_l)(k_l.k_m)(k_m.k_i)
\label{eq12}
\end{eqnarray}

Defining
$s=(k_1+k_2)^2, t=(k_1-k_3)^2, u=(k_1-k_4)^2$ 
and 
$s^\prime=(k_3+k_4)^2, t^\prime=(k_2-k_4)^2,u^\prime=(k_2-k_3)^2$,  
we can write 
$k_1.k_2=s/2, k_3.k_4=s^\prime/2, k_1.k_3=-t/2, 
k_2.k_4=-t^\prime/2, k_1.k_4=-u/2, k_2.k_3=-u^\prime/2$

Eq.~\ref{eq9} contains twelve terms. Using Eqs.~\ref{eq10},~\ref{eq11}
and Eq.~\ref{eq12} the first two 
terms of Eq.~\ref{eq9} can be expressed  as:
\be
\frac{1}{2×}g^6\frac{N_c^3}{N_c^2-1×}
\frac{N}{k_1.k_3k_1.k_4k_2.k_4k_2.k_5k_3.k_5×}
\label{eq13}
\ee
and
\be
\frac{1}{2×}g^6\frac{N_c^3}{N_c^2-1×}
\frac{N}{k_1.k_3k_1.k_5k_2.k_4k_2.k_5k_3.k_4×}
\label{eq14}
\ee
respectively. All other terms in Eq.~\ref{eq9} can be reduced in the
above form by following similar procedure.

The quantity $N$ can be written as: \\
\begin{eqnarray}
N=\frac{1}{16×}(s^4+t^4+u^4+s\prime^4+t\prime^4+u\prime^4)\nonumber\\
+\sum_{i=1}^{4}(k_i.k_5)^4
\label{eq15}
\end{eqnarray}
In the infrared and small angle scattering limits, we have :
$k_5 \rightarrow 0, t^\prime \rightarrow t, 
s^\prime \rightarrow s, u^\prime \rightarrow u$,
$s \rightarrow -u$. 

Using these approximations we get,
\begin{eqnarray}
\left| M_{gg \rightarrow ggg} \right|^2 =
g^6\frac{N_c^3}{N_c^2-1×}s^4\lbrack\frac{1}{st^2k_2.k_5k_3.k_5×}+  \nonumber  \\  \frac{1}{st^2k_1.k_5k_2.k_5×}+ 
\frac{1}{st^2k_1.k_5k_4.k_5×}+      \frac{1}{st^2k_3.k_5k_4.k_5×}\rbrack+    \nonumber  \\
g^6\frac{N_c^3}{N_c^2-1×}s^4\lbrack\frac{1}{s^3k_1.k_5k_2.k_5×}-  \frac{1}{s^2tk_2.k_5k_4.k_5×}- \nonumber  \\
\frac{1}{s^2tk_1.k_5k_3.k_5×}+\frac{1}{s^3k_2.k_5k_3.k_5×}-  \frac{1}{s^2tk_1.k_5k_3.k_5×}+  \nonumber  \\
\frac{1}{s^3k_4.k_5k_3.k_5×}+\frac{1}{s^3k_1.k_5k_4.k_5×}\nonumber\\
-\frac{1}{s^2tk_2.k_5k_4.k_5×} \rbrack 
\label{eq16}
\end{eqnarray}

The above equation may be simplified to obtain:
\begin{eqnarray}
\left| M_{gg \rightarrow ggg} \right|^2=4g^6\frac{N_c^3}{N_c^2-1×}\frac{s^3}{t^2×}\frac{1}{k_1.k_5k_2.k_5×}+ \nonumber \\
2g^6\frac{N_c^3}{N_c^2-1×}\lbrack
\frac{2s}{k_1.k_5k_2.k_5×}-\frac{s^2}{t(k_2.k_5)^2×}-
\frac{s^2}{t(k_1.k_5)^2×}\rbrack
\label{eq17}
\end{eqnarray}

The first term of the above equation is given by
\begin{eqnarray}
 \left| M_{gg \rightarrow ggg} \right|_{1st}^2=4g^4\frac{N_c^2}{N_c^2-1×}\frac{s^2}{t^2×}
\frac{g^2N_cs}{k_1.k_5k_2.k_5×} \nonumber \\ 
= \left| M_{gg \rightarrow gg} \right|^2\frac{g^2N_cs}{k_1.k_5k_2.k_5×}
\label{eq18}
\end{eqnarray}

Using the following relations~\cite{wong}, 
\be
k_\perp^2=4k_1.k_5k_2.k_5/s,
\label{eq19}
\ee
\be
q_\perp^2=4k_1.k_4k_2.k_4/s,
\label{eq20}
\ee
and
\be
(k_\perp-q_\perp)^2=4k_1.k_3k_2.k_3/s,
\label{eq21}
\ee
We get,
\be
\frac{4g^2N_cq_\perp^2}{k_\perp^2\lbrack(k_\perp-q_\perp)^2+m_D^2\rbrack}=
\frac{g^2N_cs}{k_1.k_5k_2.k_5×}
\label{eq22}
\ee
and substituting Eq.\ref{eq22} in Eq.~\ref{eq18} we obtain, 
\begin{eqnarray}
\left| M_{gg \rightarrow ggg} \right|_{1st}^2=\left(\frac{4g^4N_c^2}{N_c^2-1×}
\frac{s^2}{(q_\perp^2+m_D^2)^2×}\right) \nonumber \\
\left(\frac{4g^2N_cq_\perp^2}{k_\perp^2
\lbrack(k_\perp-q_\perp)^2+m_D^2\rbrack}\right)
\label{eq23}
\end{eqnarray}
This is the result reported in Ref.~\cite{wong}. Now we focus on the 
correction  to this result
i.e the remaining term in $\left| M_{gg \rightarrow ggg} \right|^2$, 
which is given by
\begin{eqnarray}
\left| M_{gg \rightarrow ggg} \right|^2_{2nd}=2g^6\frac{N_c^3}{N_c^2-1×}\lbrack\frac{2s}{k_1.k_5k_2.k_5×}- \nonumber \\
\frac{s^2}{t(k_2.k_5)^2×}-\frac{s^2}{t(k_1.k_5)^2×}\rbrack
\label{eq24}
\end{eqnarray}
Using Eq.~\ref{eq19} we have
\begin{eqnarray}
\left| M_{gg \rightarrow ggg} \right|^2_{2nd}=2g^6\frac{N_c^3}{N_c^2-1×}\lbrack\frac{2s}{k_1.k_5k_2.k_5×} \nonumber \\
-\frac{16(k_1.k_5)^2}{tk_\perp^4×}-\frac{16(k_2.k_5)^2}{tk_\perp^4×}\rbrack
\label{eq25}
\end{eqnarray}
In the limit  $k_5 \rightarrow 0$, we have
\begin{eqnarray}
\left| M_{gg \rightarrow ggg} \right|^2_{2nd}=2g^6\frac{N_c^3}{N_c^2-1×}\lbrack\frac{2s}{k_1.k_5k_2.k_5×}\rbrack \nonumber \\
=\frac{g^2N_cs}{k_1.k_5k_2.k_5×}\frac{4g^4N_c^2}{N_c^2-1×} \nonumber \\
=\frac{16g^6N_c^3}{N_c^2-1×}\frac{q_\perp^2}{k_\perp^2\lbrack(k_\perp-q_\perp)^2
+m_D^2\rbrack}
\label{eq26}
\end{eqnarray}
This is the correction term to the results obtained in 
~\cite{wong,GB} for the process $gg\rightarrow 2+g$.

{\bf Acknowledgment:} 
We are grateful to Raju Venugopalan for very useful discussions. 
JA would like to thank Brookhaven National Laboratory for hospitality 
when this topic was discussed. This work is supported by DAE-BRNS project 
Sanction No.  2005/21/5-BRNS/2455.

\end{document}